\documentclass[12pt]{article}
\usepackage{amsmath}
\usepackage{amssymb}
\usepackage[dvips]{graphicx}
\usepackage{color}

\newcommand {\beq}{\begin{eqnarray}}
\newcommand {\eeq}{\end{eqnarray}}

\makeatletter

  \@addtoreset{equation}{section}
\makeatother

\makeatletter

\def\mn{\def\nonumber{\global\@eqnswtrue}}
\makeatother

\makeatother

\allowdisplaybreaks[1]

\begin{document}

\begin{titlepage}
\begin{flushright}
RIKEN-MP-97\\
CALT-TH-2014-165
\end{flushright}

\begin{center}
 
\hfill 
\vskip 0.4in

{\large Is cosmological constant screened in Liouville gravity with matter?}\\[14mm] 

{Takeo Inami${\,}^a$, Yoji Koyama${\,}^b$, Yu Nakayama${\,}^c$ and Mariko Suzuki${\,}^d$}
\vskip 14mm
${}^a\,${\itshape Mathematical Physics Lab., Riken Nishina Center, Saitama, Japan}\\[1mm]
${}^a\,${\itshape Department of Physics, National Taiwan University, Taipei, Taiwan, R.O.C.}\\[1mm]
${}^b\,${\itshape Department of Physics, National Tsing-Hua University, Hsinchu, Taiwan, R.O.C.}\\[1mm]
${}^c\,${\itshape California Institute of Technology, 452-48, Pasadena, California, USA}\\[1mm]
${}^d\,${\itshape Department of Physics, Shizuoka University, 836 Ohya, Suruga-ku, Shizuoka, Japan}\\[1mm]
${}^d\,${\itshape Graduate School of Science and Technology, Shizuoka University 3-5-1 Johoku, Naka-ku Hamamatsu-shi, Shizuoka, Japan}\\

\vskip 14mm 
\end{center}
\begin{abstract}
There has been a proposal that infrared quantum effects of massless interacting field theories in de-Sitter space may provide time-dependent screening of the cosmological constant. As a concrete model of the proposal, we study 
the three loop corrections to the energy-momentum tensor of massless $\lambda \phi^4$ theory
in the background of classical Liouville gravity in $D=2$ dimensional de-Sitter space. 
We find that the cosmological constant is screened in sharp contrast to the massless $\lambda \phi^4$ theory in $D=4$ dimensions due to the sign difference between the cosmological constant of the Liouville gravity and that of the Einstein gravity. 
To argue for the robustness of our prediction, we introduce the concept of time-dependent infrared counter-terms and examine if they recover the de-Sitter invariance in the $\lambda \phi^4$ theory in comparison with the Sine-Gordon model where it was possible.

\end{abstract}

\end{titlepage}

\parskip=0.5em
\tableofcontents
\parskip=0.25em
\newpage

\section{Introduction}

Recent observation of dark energy in our universe has led to the conviction that there exists a tiny but positive value of the cosmological constant $\Lambda$. It means that our space-time is de Sitter (dS) space with the Hubble constant $H$ being $\sqrt{\Lambda}$. There has been a proposal that the strong infrared (IR) divergence property of the quantum corrections on dS space may explain the smallness of $\Lambda$ in our current universe  (so called cosmological constant problem). If we are to calculate quantum corrections to the value of the cosmological constant today, we have to deal with quantum field theories on dS space. This can be performed by using the in-in formalism or Schwinger-Keldysh formalism \cite{Schwinger:1960qe,Keldysh:1964ud,Chou:1984es,Calzetta:1986ey,Jordan:1986ug,Weinberg:2005vy}.  

The IR divergence makes it difficult to keep the dS invariance in the propagators of massless fields. The question of whether or not to preserve the dS invariance has been addressed repeatedly \cite{Ford:1977in,Allen:1987tz,Allen:1986dd,Allen:1986tt,Kleppe:1993fz,Miao:2010vs,Rajaraman:2010zx,Higuchi:2011vw,Hollands:2011we,Giddings:2011ze,Miao:2012xc,Youssef:2012cx,Morrison:2013rqa} (for IR effects during cosmological inflation, see \cite{Seery:2010kh} for a review). The ambiguity in imposing the boundary condition on the propagators at the horizon has also been discussed in \cite{Urakawa:2009my}. A complete agreement has not been reached yet in the evaluation of loop corrections to the energy-momentum (EM) tensor $T_{\mu\nu}$ in quantum gravity coupled to matter in the four dimensions (4D) \cite{Tsamis:1996qq,Garriga:2007zk,Kitamoto:2012ep,Kitamoto:2014gva}. In this situation we believe that studying quantum gravity and matter loop effects on the EM tensor $T_{\mu\nu}$ in two-dimensional (2D) dS space may help clarify the problem of quantum corrections to the cosmological constant $\Lambda$.

Our $D=2$ dimensional model for quantum gravity coupled to matter fields is based on the 2D Liouville field theory \cite{Nakayama:2004vk} minimally coupled to matter fields. The Liouville field is a Weyl factor of the metric and originally it has no kinetic term at the classical level.
The origin of the kinetic term is from the Weyl non-invariant measure of the path integral for quantum gravity such as Weyl anomaly. The resulting Liouville field theory captures the non-perturbative dynamics of the low energy effective field theory of the 2D quantum gravity \cite{Polyakov:1987zb,Distler:1988jt} and hence contains the complete information of the quantum gravity as an ordinary quantum field theory. Once we derive the Liouville field theory as the 2D quantum gravity (Liouville gravity), one may take the classical limit by assuming the large number of matter fields. 
The Liouville field theory is conformally invariant, and at least classically, there is no subtleties in the dS background.
The ``coupling constant" of the potential term in the Liouville field theory is related to the cosmological constant of 2D quantum gravity and renormalized by the matter loop effects.   
Similarly to the higher-dimensional Einstein equation with the dS breaking quantum matter EM tensor, we expect that the subtle quantum IR effects of the interacting massless fields may significantly affect the dynamics of the  Liouville gravity. 

Constructing a 2D model based on the Liouville field theory is also motivated by (and is related to) an old work by Polyakov for the IR screening of the cosmological constant \cite{Polyakov:1982ug}. The Weyl factor of the metric plays a leading role there. 
There are attempts to the screening mechanism from the dynamics of the Weyl factor in 4D gravity \cite{Antoniadis:1991fa,Jackiw:2005yc,Antoniadis:2006wq} and, more relevant to our work, in 2D dilation gravity \cite{Govaerts:2011ig} (including Liouville gravity as a special case) where the cosmological constant is canonically quantized non-perturbatively. If such a mechanism is really at work in de-Sitter space-time, it would significantly affect our mind-set to understand the cosmological constant problem. 
Since the Liouville gravity coupled to quantum matter is a power-counting renormalizable
field theory, we should be able to answer the question unambiguously.\footnote{In \cite{Kitamoto:2014gva}, a possibility to screen the cosmological constant in non-unitary time-like Liouville theory was discussed. See also footnote 5 in comparison with their approach.} To be screened or not to be screened, that is the question.

In this work we restrict ourselves to the perturbative effects from the matter sector. As a concrete example, we choose a scalar field theory with $ \lambda \phi^4$ interaction on dS background.  
Evaluation of the matter loop corrections to $T_{\mu\nu}$ to higher loops is carried out by using the propagator for a massless scalar field $\phi$ with a dS symmetry breaking term. Hence the EM tensor acquires the logarithmic time dependence which is often referred to as the IR logarithm, $\ln a(t)$. We find that our result shows the screening effect of the effective cosmological constant up to order $\lambda^3$ corrections. We also find that the degree of IR divergence from massless scalar fields in 2D dS space is the same as that in 4D at least within a perturbative computation.  

This conclusion, however, is puzzling at least for one reason. In flat Minkowski space, the IR limit of $ \lambda \phi^4$ theory in $D=2$ dimensions is equivalent to a free massless Majorana fermion (or critical Ising model) from the Landau-Ginzburg construction \cite{DiFrancesco:1997nk,Zamolodchikov:1986db}. The free Majorana fermion is conformally invariant and does not show any IR pathology in dS space. The cosmological constant induced by the free Majorana fermion is never screened.

A similar puzzling situation was reported in the literature \cite{Bander:2010pn} in relation to the  (in)equivalence between the Sine-Gordon model and massive Thirring model in dS background. If we quantize the Sine-Gordon model perturbatively around the massless scalar field theory in dS space, the quantum IR effect of the  massless scalar propagator appears as in $\lambda\phi^4$ theory and the dS symmetry is broken. On the other hand, in the dual fermionic picture, there is no such breaking effect at all. In this particular case, however, we will see in the last section that by adding non-conventional dS breaking local counter-terms (with which the quantum dS breaking is cancelled in the final correlation functions), we may recover the dS invariance in Sine-Gordon model. 

Whether such counter-terms are allowed or should be added must be determined from some other principles. If we stick to the dS invariance, there is no reason not to add them unless it is inconsistent with more important principles such as gauge invariance. The mechanism should work in other space-time dimensions while at this stage, we are not certain if the obstruction to recover the dS invariance from IR counter-terms existed for gravitons or gauge fields in higher dimensions.

To argue for the robustness of our prediction, we explore such a possibility in $\lambda \phi^4$ theory as well. Certainly, we may cancel the dS breaking screening effects on cosmological constant by adding the time-dependent classical IR counter-term by hand. However, unlike Sine-Gordon model, we do not find systematic ways to remove the dS breaking effects in the other correlation functions by time-dependent but local counter-terms. So within our perturbation theory, the quantum IR effects in $\lambda \phi^4$ theory are real phenomena and it is not natural to cancel only the time-dependence in the cosmological constant.

In the next section, we briefly review the IR divergence originated from non-conformally invariant massless scalar fields in dS space in general space-time dimensions. Introduction of an IR cutoff for momentum integration leads to the IR logarithm $\ln a$ in the coordinate space propagator which immediately breaks a part of dS symmetry, namely dilatation invariance, $\eta \to b \eta$, $x^{i}\to bx^{i}$. It makes the cosmological constant time dependent through the Einstein equation. We introduce our 2D model of Liouville gravity and matter loop corrections in section 3. In section 4, we compute the perturbative corrections of order $\lambda^2$ to the EM tensor from massless scalar loops in $ \lambda \phi^4$ theory. In section 5, we discuss the possibility of the dS non-invariant counter-terms designed for cancelling the dS breaking IR logarithms in comparison with Sine-Gordon model in dS space.  We conclude with discussion in section 6.
In appendix A, we report the detailed computation of the order $\lambda^2$ corrections to the EM tensor in $ \lambda \phi^4$ theory.

\section{Cosmological constant problem and infrared effects}

\subsection{Infrared divergences in de-Sitter space}

In this paper, we work on quantum field theories on $D$-dimensional dS space. Among various choices of coordinates, we mainly use the  Poincar\'{e} coordinate, in which it is manifested that dS space is conformally flat.
The dS geometry is expressed by the metric
\beq
ds^2= -dt^2 +a^2(t)d{\vec x}\cdot d{\vec x},
\eeq
where the scale factor $a$ is given by Hubble constant $H$, and the conformal time $\eta$ as
\beq
a =e^{Ht}=-\frac{1}{H\eta}, \,\, H(t)\equiv \frac{\dot{a}(t)}{a(t)}.
\eeq
Here $\eta$ is related to the physical time $t$ by
\beq
\eta = - \frac{1}{H} e^{-Ht},
\eeq
and it runs from $-\infty$ to 0 ($-\infty\leq t \leq \infty$). By using the conformal time $\eta$, the dS metric becomes
\beq
ds^2= a^2(\eta)(-d\eta^2 +d{\vec x}\cdot d{\vec x} ).
\eeq
This coordinate covers the half of the global dS space.

In dS space, IR divergence property of (non-conformally invariant) massless fields is different from that in Minkowski space because large distance is affected by the dS curvature. One can easily see such a property by considering the vacuum loop graphs of massless scalar fields.  They are obtained by integrating over the loop momentum $P$ where $P$ is physical momentum. Let us follow the explanation given in \cite{Kitamoto:2011yx}.
It is convenient to divide the integration region into two, UV region (sub-horizon) $|P|>H$ and IR region (super-horizon) $|P|<H$. For example in 4D space-time\footnote{In terms of comoving momentum, we have
$\int d^3k = \int_{|k|>aH} d^3k +\int_{|k|<aH} d^3k$.
The IR cutoff is given by $k_{0}=a_iH$ \cite{Urakawa:2009my} where $a_i$ is the scale factor at the initial time. } 
\beq
\int d^3P = \int_{|P|>H} d^3P +\int_{|P|<H} d^3P.
\label{region}
\eeq
The mode function of massless minimally coupled scalar field in the Bunch-Davies vacuum \cite{Bunch:1978yq} is given by
\beq
\phi^{D=4}_{\vec k}(\eta)=\frac{H\eta}{\sqrt{2k}}
\left( 1-\frac{i}{k\eta} \right)
e^{-i k\eta} ,\label{mode}
\eeq
where $k$ is a comoving momentum which is related to physical momentum $P$ as $P=k/a(\eta)=-k H\eta$. In dS space the fluctuations of the massless fields (scalars and gravitons) have the scale free spectrum which behaves as $1/P^3$ at super-horizon scale due to the second term in (\ref{mode}). The loop integral $\int_{|P|<H} d^3PP^{-3}$ then gives rise to a logarithmic divergent contribution at $P\to 0$.

Let us repeat the analysis in $D=2$, which is our main focus of this paper. The mode function of the massless scalar field is given by
\beq
\phi^{D=2}_{\vec k}(\eta)=\frac{-i}{\sqrt{2k}}
e^{-i k\eta}.
\eeq
This is completely the same form of that in Minkowski space because minimally coupled free massless scalar fields in 2D space-time are conformally invariant and we have a conformal vacuum as a dS invariant vacuum. The vacuum loop is in this case given by
\beq
\int^{\infty}_{0}dP \frac{1}{2P}=\int^{\infty}_{H}dP \frac{1}{2P}+\int^{H}_{0}dP \frac{1}{2P},\label{2Dintegral}
\eeq
where we again make use of physical momentum $P=k/a(\eta)$ and divide the integral into $P<H$ and $P>H$ as in $D=4$ case even though there is no  distinction between the behavior of sub-horizon and super-horizon modes. The divergence structure is the same as in $D=2$ Minkowski space.

As we have seen in above two examples, the origin of the IR divergence lies at (i) zero comoving momentum $k\to0$ or (ii) infinite future $\eta\to 0$. Here we regularize the IR divergence from (i) by truncating the Hilbert space at some comoving momentum $k_0$ as an IR cutoff. As a result of this prescription, the second term in (\ref{region}), and (\ref{2Dintegral}) give a factor $\ln (a(\eta)/k_0).$\footnote{
For later use we note that the IR cutoff $k_0$ is related to the cutoff for the initial time $t_i$ \cite{Kitamoto:2011yx,Urakawa:2009my} since the largest comoving scale is given by $l_0=k_0^{-1}$, and an identification of $l_0$ leads to that of $t_i$ through $a(t_i)l_0=L_i$ with $L_i$ the initial (physical) size of the universe. We adopt $a(t_i)=1\ (t_i=0)$ with $L_i=l_0=H^{-1}$ as a reference time for the renormalization conditions in section 4 and an initial time cutoff for the vertex integrals in Appendix A.}  

There is a little subtlety in putting IR and UV cutoffs for momentum in the vacuum loop. To obtain the IR logarithm from the vacuum loop, the UV cutoff of the first integral in (\ref{region}) has been taken implicitly to be {\it physical} $P=\Lambda_{\rm UV}=\mathrm{const.}$, which makes 
\beq
\int^{\Lambda_{\rm UV}}_{H}dP\notag
\eeq
constant. On the other hand, the IR cutoff in the second integral is taken to be {\it comoving}, $k_0$, and accordingly, $P=k_0/a(t)$ is not a constant. It amounts to saying that UV cutoff of the theory does not change due to the cosmic expansion, on the other hand, the number of IR modes ($P(t)<H$) increases with time. If we put the IR cutoff to be physical $P=\Lambda_{\rm IR}=\mathrm{const.}$, the time dependence disappears from the vacuum loop.   
   
Even if the IR divergence is regularized once, owing to the exponential expansion of the universe, the dS space distance grows with time and at the same time the physical wavelengths are all red shifted. Eventually as the conformal time $\eta$ approaches $0$ ($t$ to $\infty$) IR divergence of kind (ii) appears due to $\ln (a(\eta))$ which is often referred to as the (dS breaking) IR logarithm. From the detailed study of this IR logarithm with dS breaking, we can learn the characteristic effect of the massless fields and its consequence in dS space. 
 
The same type of IR divergence as in the case of the vacuum loop appears in the coordinate space propagator. In section 4, we will adopt the dimensional regularization for UV divergence in the loop computations while we adopt the cutoff regularization for IR divergence. The propagator of a massless minimally coupled scalar field in $D$-dimensional dS space is obtained in \cite{Onemli:2002hr,Janssen:2008px}. By setting $D=2-\omega$ with $\omega>0$, the propagator is given by
\beq
i\Delta(x,z)=\alpha \{ \gamma(y)+\beta \ln(a(\eta)a(\eta_z))\},
\label{propagator}
\eeq
where
\beq
y&\equiv&\frac{-(\eta-\eta_z)^2+({\vec x}-{\vec z})^2}{\eta \eta_z},
\eeq
\beq
\alpha&=&\frac{1}{4\pi}\left(\frac{H^2}{4\pi}\right)^{-\omega},\quad \beta=\frac{\Gamma(1-\omega)}{\Gamma(1-\frac{\omega}{2})},
\eeq
and
\beq
\hspace{-0.8cm}\gamma(y)&=&-\frac{\Gamma(1-\frac{\omega}{2})}{\omega}\left(\frac{y}{4}\right)^{\frac{\omega}{2}}-\frac{\Gamma(2-\frac{\omega}{2})}{1+\frac{\omega}{2}}\left(\frac{y}{4}\right)^{1+\frac{\omega}{2}}+\beta\delta\nonumber\\
&&+\sum^{\infty}_{n=1} \left[ \frac{\Gamma(1-\omega+n)}{n\Gamma(1-\frac{\omega}{2}+n)} -\frac{\Gamma(2-\frac{\omega}{2}+n)}{(1+\frac{\omega}{2}+n)(n+1)!}\left(\frac{y}{4}\right)^{\frac{2+\omega}{2}}\right]\left(\frac{y}{4}\right)^n +\mathcal{O}(k_0^2),
\label{pros}
\eeq
with
\beq
\delta&\equiv&-\pi\cot\left(\pi-\frac{\omega}{2}\pi\right)+C,\\
C&\equiv& \frac{1}{2}\ln\left(\frac{H}{k_0}\right)+\psi\left(\frac{1}{2}-\frac{\omega}{2}\right)-\psi\left(1-\frac{\omega}{2}\right)+\psi(1-\omega)-\gamma.
\label{proe}
\eeq
Here an IR cutoff $k_0$ for comoving momentum has been introduced. $\psi(x)$ is the digamma function and $\gamma$ the Euler-Mascheroni constant. The distance in dS space is commonly denoted by $y$. The propagator (\ref{propagator}) has a simple structure that the first term is manifestly dS invariant because it 
depends only on the distance $y$ which respects the dS symmetry in Poincar\'{e} coordinate (a dilatation, $D-1$ dimensional spatial rotations, $D-1$ dimensional spatial translations and $D-1$ dimensional special conformal transformations \cite{Kleppe:1993fz,Miao:2010vs,Bander:2010pn}). On the other hand, the second term breaks the dS invariance (the dS isometry $\eta \to b \eta$ and $x \to b x$ especially) due to the the IR logarithm. 

The basic formalism to calculate the correlation functions in time dependent backgrounds is called the in-in formalism. In the in-in formalism two copies of time sheets, named by $+$ and $-$ are prepared and the time path is then closed: $\int_C d\eta=\int_{-\infty}^0 d\eta_++\int^{-\infty}_0 d\eta_-$. All vertices in the loop diagrams are assigned $+$ or $-$ type. The expectation value of operator(s) ${\cal O}(x)$ is given by
\beq
\langle\Omega| {\cal O}(x)|\Omega\rangle
&=&{\langle {T}\{{\cal O}(x)e^{-i\int_C H_{{\rm int}}d\eta}\} \rangle}\notag\\
&=&{\langle  {\tilde T}\{e^{i\int_{\eta_i}^0 H_{{\rm int}-}d\eta_-}\}
{T}\{{\cal O}(x)e^{-i\int_{\eta_i}^0 H_{{\rm int}+}d\eta_+}\} \rangle}
 \label{vevinin}
\eeq
 where $T$ and ${\tilde T}$ stand for the usual time ordering operator and the anti-time ordering operator, respectively.
Here $|\Omega\rangle$ in the first line is the vacuum of the interacting theory, and the $\langle \mathcal{O}(x) \cdots \rangle$ is the expectation value in the free field theory that can be computed by the Wick contraction.
We introduced $\eta_i$ as an initial time and assumed $x$ to be $+$ type in the second equality. Depending on the types of vertices, all $y$ have one of four types \cite{Onemli:2002hr}.
\beq
\begin{split}
y_{++}(x,x') &\equiv a(\eta)a(\eta ') H^2 [({\vec x}-{\vec x}')^2 -(|\eta-\eta'|-ie)^2],\\
y_{+-}(x,x') &\equiv a(\eta)a(\eta ') H^2 [({\vec x}-{\vec x}')^2 -(\eta-\eta'+ie)^2],\\
y_{-+}(x,x') &\equiv a(\eta)a(\eta ') H^2 [({\vec x}-{\vec x}')^2 -(\eta-\eta'-ie)^2],\\
y_{--}(x,x') &\equiv a(\eta)a(\eta ') H^2 [({\vec x}-{\vec x}')^2 -(|\eta-\eta'|+ie)^2],
\end{split}
\eeq
where $y_{ab}(x,x')$ stands for $y(x_a,x'_b)$ ($a,b=\pm$) with $e$ a positive infinitesimal.
By substituting each distance $y$, we can construct the four propagators used in the in-in formalism. We denote them by
\beq
\begin{split}
 i\Delta_{++}(y_{++})&\equiv\langle T\{\phi(x_+)\phi(x_+')\}\rangle,\\ 
 i\Delta_{+-}(y_{+-})&\equiv\langle \phi(x_-')\phi(x_+)\rangle,\\
 i\Delta_{-+}(y_{-+})&\equiv\langle \phi(x_-)\phi(x_+')\rangle,\\
i\Delta_{--}(y_{--})&\equiv\langle {\tilde T}\{\phi(x_-)\phi(x_-')\}\rangle.\label{propagators}
\end{split}
 \eeq
The short distance $(y \to 0)$ limit of the propagator (\ref{propagator}) is regularized by $\omega$ and is independent of the labels $+$, $-$, as is seen
\beq
  \lim_{x \to z } i\Delta(x,z)
=\alpha\beta \left( 2\ln(a(\eta))+\frac{2}{\omega}+C+\gamma+\mathcal{O}(\omega)\right).
\label{coincidence}
\eeq

\subsection{Infrared effects on the cosmological constant $\Lambda$}

In space-time dimensions $D>2$, the Einstein equation describes the relation between the space-time Ricci tensor $R_{\mu\nu}$ and the EM tensor $T_{\mu\nu}$ due to the presence of matters,
\beq
R_{\mu\nu} - \frac{1}{2}g_{\mu\nu}R+\Lambda g_{\mu\nu} =\kappa T_{\mu\nu},\label{einsteineq}
\eeq
where $R$ is scalar curvature, $\kappa=8\pi G$ with $G$ being Newton's constant, $\Lambda$ is the cosmological constant. In the vacuum states where $T_{\mu\nu}$ is proportional to the metric, we may transfer $T_{\mu\nu}$ to the left hand side of (\ref{einsteineq}), and the vacuum Einstein equation takes the form
\beq
R_{\mu\nu} - \frac{1}{2}g_{\mu\nu}R+\Lambda _{\rm{eff}}g_{\mu\nu}=0,
\eeq
where
\beq
\Lambda_{\rm{eff}}=\Lambda -\frac{\kappa}{D}T_\rho ^{\,\,\,\rho}.
\label{EMTeff}
\eeq
The vacuum contribution of $T_{\mu\nu}$ is now combined with $\Lambda$ to define the effective cosmological constant. 

In view of this expression we may wonder if a large value of $T_{\mu\nu}$ may cancel the large value of $\Lambda$ yielding a tiny value of $\Lambda_{\rm{eff}}$ that we observe today. For such cancellation, the quantum corrections to $T_{\mu\nu}$ is essential. This idea may or may not address the cosmological constant problem because we have yet to know what the bare cosmological constant $\Lambda$ should be (see \cite{Polchinski:2006gy} for a review).

The situations in dS space with massless interacting fields are much more complicated. As mentioned previously, the massless scalar propagator in dS space has IR divergence. This IR divergence is regulated by the IR cutoff and  renormalized.
The IR cutoff, however, introduces dS invariance breaking term from the IR logarithm $\ln a$. Then the expectation value of $T_{\mu\nu}$ of the massless interacting fields in dS space explicitly depends on the IR logarithm $\ln a$ and it becomes non-dS-covariant. As a consequence the effective cosmological constant becomes time-dependent from \eqref{EMTeff}. This time-dependent screening effects proposed in the literature may cause the drastic effects in the fine-tuning problem of the cosmological constant.

In perturbation theory, both matter loops and gravity loops may provide sources of corrections to the cosmological constant $\Lambda$. Quantum effects of gravity in 4D dS space have been studied extensively for a long time (see \cite{Woodard:2014jba,Klemm:2004mb} for reviews).
Due to the difficulty of keeping the dS invariance in the massless propagators and the ambiguity in taking account of the boundary conditions at the horizon, a  complete agreement has yet to be reached in the evaluation of loop effects on the $T_{\mu\nu}$ even after extensive studies. In the semi-classical limit, or in the large number of matter fields limit, the matter loop corrections will dominate over the gravity loop corrections, so we may treat the Einstein gravity classically while replacing $\Lambda$ by the quantum expectation value of the matter contributions in the fixed dS background.
Although such a limit is purely academic in our $D=4$ universe, we may still learn important lessons on screening of cosmological constant from the quantum IR effects.

The goal of this paper is to calculate quantum effects in lower dimension because IR divergence in lower dimension is stronger than that in higher dimension in Minkowski space.  A question is whether we observe similar enhancement of the IR effects in $D=2$ dS space.
We also address the question if the dS invariance may be recovered from the IR counter-terms. While we demonstrate the possibility in $D=2$, the same argument may apply in higher dimensions, too.

In $D=2$, the classical Einstein gravity becomes trivial, and the discussion in this section must be replaced by the other model of gravity. We opt to use the Liouville gravity that is induced by the quantum fluctuation of the Weyl mode of the metric. Again in the large number of matter fields limit, one may treat the Liouville degrees of freedom classically while replacing the effective cosmological constant term from the matter contributions in the fixed Liouville background.
The details will be described in the next section.

\section{2D model for quantum gravity with matter}

\subsection{2D Liouville theory}

In $D=2$ dimensions, the Einstein gravity with the Einstein-Hilbert action
\beq
S[g_{\mu\nu}]=\int d^2 x \sqrt{-g} \frac{1}{2\kappa}\left(R-2\Lambda \right),\label{EHaction}
\eeq
has no dynamical degrees of freedom because the Einstein-Hilbert term is topological due to Gauss-Bonnet theorem. 
However, at the quantum level, the Weyl mode $\Phi$ of the metric $g_{\mu\nu} = e^{2\Phi}\hat{g}_{\mu\nu}$ becomes dynamical and the {\it quantum} gravity in $D=2$ dimensions is described by the dynamical Liouville field theory. Here $\hat{g}_{\mu\nu}$ is the fiducial metric that we can choose arbitrarily. Because of this arbitrariness, the Weyl invariance (i.e. $\hat{g}_{\mu\nu} \to e^{2\sigma} \hat{g}_{\mu\nu}$ and $\Phi \to \Phi -\sigma$ ) of the Liouville gravity is automatically guaranteed.\footnote{More precisely, the quantization must respect the symmetry.}

 We briefly recapitulate the Liouville theory. We begin with 2D gravity field (metric) $g_{\mu\nu}$ coupled to ``matter fields" collectively called $X$. The action and the partition function are given by 
\beq
S_{\mathrm{2D}}[g_{\mu\nu},X]&=&\int d^2x \sqrt{-g}\left(\frac{1}{2\kappa}(R- 2\Lambda) \right) + S_{\mathrm{matter}}[g_{\mu\nu},X] ,\\
Z&=&\int D X Dg_{\mu\nu}e^{iS_{{\mathrm{2D}}}[g_{\mu\nu},X]} \ .
\eeq
In $D=2$, we may (locally) parametrize the gravity fluctuation by the Liouville degrees of freedom $g_{\mu\nu} = e^{2\Phi} \hat{g}_{\mu\nu}$ with the fiducial metric $\hat{g}_{\mu\nu}$. In this conformal gauge, the path integral over $g_{\mu\nu}$ is replaced by the path integral over the Liouville field $\Phi$ with the appropriate measure factor. Since the Einstein-Hilbert term only gives the topological contribution, we drop $\frac{R}{2\kappa}$ term in the following.

The path integral measure of the 2D quantum gravity also contains the diffeomorphism ghost factors in the conformal gauge, but we will ignore them for our purpose since it has little to do with our interest in the geometric dynamics of the Liouville field (except for the balance of the Weyl anomaly). As for the measure factors of the Liouville field, it is expected to be ultra local and gives the kinetic term of the Liouville field. We note that the kinetic term of the Liouville action is also induced by the Weyl anomaly
\beq
\langle T_{\ \ \ \  \rho} ^{\text{mat} \ \rho}\rangle = \frac{c_{\mathrm{matter}}}{24\pi} R,\label{Liouvilletrace}
\eeq
of the matter action, where $c_{\mathrm{matter}}$ is the matter central charge. Indeed, the Liouville action may be regarded as the local Wess-Zumino like term for the Weyl anomaly.

Collecting all these quantum contributions to the Liouville degree of freedom, the matter-gravity action can be reformulated as
\beq
Z\sim \int D\Phi DX e^{iS_{L}(\Phi) + i S_{\mathrm{matter}}[\Phi,X] },
\eeq
where 
\beq
S_{L}=-\int d^2 x \sqrt{-\hat{g}}\left( \frac{1}{4\pi b^2 }\hat{g}^{\mu\nu}\partial_\mu \Phi \partial_\nu \Phi +\frac{Q}{4\pi b} \Phi R(\hat{g})+\frac{\Lambda_{\mathrm{ren}}}{\kappa}e^{2\Phi}\right). \label{lga}
\eeq
Here, $\Lambda_{\mathrm{ren}}$ is the renormalized cosmological constant.
We will drop the subscript $\mathrm{ren}$ in the following.
 $Q$ is the background charge given by $Q= b + b^{-1}$  (see e.g. \cite{Nakayama:2004vk}) so that the total action is quantum mechanically conformally invariant. In the classical limit $(b \to 0)$ that we will discuss below, we have the value $Q=b^{-1}$.

If the matter action is conformally invariant, the Liouville field does not appear in the matter action $S_{\mathrm{matter}}[\Phi,X]$. We will discuss the matter coupling in the next subsection, and we focus on the Liouville part for now. The Liouville field theory is a conformal field theory in a fixed background $\hat{g}_{\mu\nu}$. 
The path integral over $\Phi$ is non-trivial, but we may use the trick of large number of matter fields  limit again. When the number of matter fields become larger, the induced Liouville kinetic term is larger and larger,\footnote{It is determined by the Weyl anomaly of the matter as $6b^{-2} \sim -c_{\mathrm{matter}}$. Actually, it is negatively smaller for larger $c_{\mathrm{matter}}$, and the kinetic term becomes negative. To avoid the difficulty, we may add the ``non-unitary" matter with the large negative central charges. For our discussions, we always keep $b^2$ to be positive. Note that this regime is opposite to the one studied in \cite{Kitamoto:2014gva} where $b^2$ was chosen to be negative.} so the quantum fluctuation of the Liouville field becomes suppressed. In \eqref{lga}, $b^2$ becomes smaller for the larger number of matter fields, and the Planck constant becomes smaller. Therefore, although the origin of the Liouville action is purely quantum mechanical, we may treat it as if it is classical in the limit of large number of matter fields. 

In analogy to the dS solution in the Einstein gravity, our interest is the dS solution of the Liouville gravity. There are two alternative viewpoints. One is to choose the background fiducial metric $\hat{g}_{\mu\nu}$  to be dS space. Then we see that the classical equations of motion of the Liouville field becomes $2\Lambda \kappa^{-1} e^{2\Phi} = -\frac{Q}{4\pi b}H^{2}$ for  constant $\Phi$. With a convenient choice of $\Phi=1$, the Hubble constant and the 2D cosmological constant (or Liouville coupling constant) is related. Note that the value of $H$ is not that important in the physical metric $g_{\mu\nu} = e^{2\Phi} \hat{g}_{\mu\nu} $ because it is cancelled by the factor $\Phi$ from the Liouville equation. At this point, it is important to remind ourselves that the negative value of $\Lambda$ corresponds to dS space in Liouville gravity (see also the discussion in the next subsection).

The other viewpoint is to consider the Liouville equation in the flat Minkowski space with $\hat{g}_{\mu\nu} = \eta_{\mu\nu}$ so that it becomes $\Box \Phi = 4\pi b^2 \Lambda \kappa^{-1} e^{2\Phi}$. The Liouville field cannot become constant, and the simplest solution is $\Phi = -\ln (-H\eta)$, which again gives rise to the physical dS metric $g_{\mu\nu} = e^{2\Phi} \eta_{\mu\nu}$. 
In whichever viewpoint, the matter action couples to the physical metric $g_{\mu\nu}$, so we may only consider the matter action in the dS space.

So far, in this section, we have treated the matter contributions as if it preserves the dS invariance. When the matter EM tensor breaks the dS invariance, the classical Liouville equation is modified and the screening effects of the Liouville coupling constant may occur. This is analogous to the matter screening of the effective cosmological constant discussed in the last section, and we will study it in the following.

\subsection{The coupling of Liouville gravity and matter}
Our main interest is to evaluate the quantum effects of gravity and matter at IR region by making use of Liouville field theory. The 2D cosmological constant has two faces, one as the coupling of the Liouville potential in terms of the Liouville action, one as the trace of the EM tensor. Let us start with the action
\begin{align}
&\hspace{-0.1cm}S_{L+\rm{mat}}[\Phi,\phi]\cr
&=-\int d^2 x \sqrt{{-g}}\left[ \frac{1}{4\pi b^2} g^{\mu\nu}\partial _\mu \Phi \partial _\nu \Phi+\frac{Q}{4\pi b} R \Phi +\frac{\Lambda}{\kappa} +\frac{1}{2}g^{\mu\nu}\partial_\mu\phi \partial_\nu \phi +V(\phi)\right].
\end{align}
After taking Weyl transformation to metric,\footnote{We have assumed that the Weyl anomaly cancels among Liouville part, matter part and the ghost part which we have not written down explicitly.} we obtain 
\begin{align}
&\hspace{-0.5cm} S_{L+\rm{mat}}[\Phi,\phi] \cr
&\hspace{-0.4cm}=-\int d^2 x \sqrt{{-\hat{g}}}\left[  \frac{1}{4\pi b^2} \hat{g}^{\mu\nu}\partial _\mu \Phi \partial_\nu \Phi+\frac{Q}{4\pi b}\hat{R}\Phi +\frac{\Lambda}{\kappa} e^{2\Phi} +\frac{1}{2}\hat{g}^{\mu\nu}\partial_\mu\phi \partial_\nu \phi +e^{2\Phi} V(\phi)\right].\label{2Dmodel}
\end{align}
 $\Phi$ is the Liouville field and $\phi$ is a matter field. The fifth term describes the interaction term between the 2D Liouville gravity and matter.

The above argument is purely classical in the Liouville degrees of freedom. As advocated before we are working in the classical Liouville regime in the large number of matter fields in mind. We will only focus on one particular degree of freedom of the matter (i.e. a scalar with $\lambda \phi^4$ interaction), but we always assume the extra large numbers of spectator matter fields to make the classical treatment of the Liouville field theory valid. 
 
 The dS symmetry plays an important role in the determination of the trace of the EM tensor together with the conformal symmetry in Liouville field theory. However we have seen in section 2 that if there is a massless scalar field, IR divergence will arise and break a part of the dS symmetry. The existence of a dS invariant vacuum then becomes ambiguous at least from the perturbative point of view. In this case we have additional time-dependent contributions to the effective cosmological constant.

As in the Einstein gravity case discussed in the previous section, the effective cosmological constant is given by
\begin{align}
\Lambda_{\mathrm{eff}} &= \Lambda + \kappa \langle V(\phi) \rangle  \cr
& = \Lambda -\frac{\kappa}{2} \langle T_{\rho}^{\ \rho}\rangle \ . \label{screening2}
\end{align}
Then the effective Liouville equation takes the form
\begin{align}
-\frac{1}{2\pi b^2} \hat{\Box} \Phi + \frac{Q}{4\pi b} \hat{R} = -2\Lambda_{\mathrm{eff}} \kappa^{-1} e^{2\Phi}  \ ,
\end{align}
where $\Lambda_{\mathrm{eff}}$ may contain the effects of the IR dS breaking from the matter contributions in \eqref{screening2}. If this is the case, the Liouville field can be no longer constant with the fiducial dS metric $\hat{g}_{\mu\nu}$. Then the physical metric $g_{\mu\nu}$ is not dS invariant in the semi-classical limit. 
In this sense, the screening of the cosmological constant gives the similar effects in the Liouville gravity to the Einstein gravity in the higher dimensions.

There is one subtle but important distinction between the Einstein gravity and the Liouville gravity that we would like to point out. In the Einstein gravity, if the energy of the universe is positive then the space-time allows the classical dS solution. This is the meaning of the positive cosmological constant in the expanding universe. However, in the Liouville gravity, the opposite is true. If the universe has the {\it negative} energy then the Liouville equation allows the classical dS solution (or sphere in the Euclidean signature). This difference yields an interesting consequence in the non-perturbative Liouville cosmology with meta-stable vacua \cite{Zamolodchikov:2006xs,Nakayama:2010ea}. In our study, the sign difference makes the IR effects of the massless $ \lambda \phi^4$ theory screen rather than anti-screen the cosmological constant in $D=2$ in sharp contrast to the situations in $D>2$.

\section{Quantum corrections --- 2D matter}

By fixing the value of the Liouville field $\Phi$ to its classical configuration, (\ref{2Dmodel}) is equal to the matter action in a fixed gravitational background. In what follows we concentrate on the dynamics of the matter field $\phi$. The purpose of this section is to evaluate the massless matter loop corrections to the EM tensor in the massless $\lambda \phi^4$ theory. The loop corrections from massless matters are an interesting problem in its own right. 
Our main interest is the IR logarithms which are particular for massless scalar fields (and graviton) in dS space. 
We are going to show explicitly that the cosmological constant indeed receives the time dependent corrections through the dS breaking expectation value of the EM tensor in the way we have discussed in section 2 and 3. 

We shall work with a 2D massless minimally coupled scalar field theory with $\lambda \phi^4$ interaction. The Lagrangian is given by
\beq
\mathcal{L}=-\frac{1}{2}g^{\mu\nu}\partial _\mu \phi \partial _\nu \phi \sqrt{-g} -\frac{1}{4!} \lambda \phi^4 \sqrt{-g} +\Delta \mathcal{L},\label{2dphi4}
\eeq
where $\Delta \mathcal{L}$ consists of the counter-terms
\beq
\Delta \mathcal{L}&=&-\frac{1}{2}\delta Z g^{\mu\nu}\partial _\mu \phi \partial _\nu \phi \sqrt{-g}-\frac{1}{4!} \delta \lambda \phi^4 \sqrt{-g} -\frac{1}{2}\delta m^2 \phi^2 \sqrt{-g}\nonumber\\
&&+\delta \xi (R-D(D-1)H^2)\phi^2\sqrt{-g}-\frac{\delta \Lambda}{8\pi G}\sqrt{-g}.
\eeq
The matter EM tensor is given by 
\beq
T^{{\rm mat}}_{\mu\nu}(x)&=&(1+\delta Z)\left(\delta^\rho_{\ \mu} \delta^\sigma_{\ \nu} -\frac{1}{2}g_{\mu\nu} g^{\rho\sigma} \right) \partial_\rho \phi \partial_\sigma \phi  - g_{\mu\nu}\left(\frac{\lambda+\delta\lambda}{4!} \phi^4 +\frac{1}{2} \delta m^2\phi^2 +\frac{\delta\Lambda}{8\pi G}\right)\nonumber\\
&\,& -2\delta\xi\left[g_{\mu\nu}((D-1)H^2 \phi^2  + (\phi ^2)^{; \rho}_{\,\,\, \rho})-(\phi^2)_{;\mu\nu}\right],
\eeq
where $;$ denotes the covariant derivative with respect to the dS background $g_{\mu\nu}=a(\eta)^2\eta_{\mu\nu}$. The first term is the effect of the kinetic term and the second, third and fourth terms are the effects of the potential terms. The last term proportional to $\delta \xi$ is the conformal counter-term.

We will calculate the vacuum expectation value (VEV) of the EM tensor by using the in-in formalism. Our calculation is regarded as a 2D analogue of that in 4D performed first in \cite{Onemli:2002hr}. Regarding the IR logarithm, $\ln a$, we expect that the leading contribution to the EM tensor comes from the potential term. It is because the degree of IR divergence is weakened by derivatives: derivatives acting on the propagators reduce the number of the IR logarithms. Hence the contributions of the IR logarithms from the kinetic terms are weaker than that from the potential terms at each order of perturbative calculation. More detailed discussions including the issue of conservation of the EM tensor are given in \cite{Kitamoto:2011yx}. In the following we will focus on the potential term as the leading contribution to the EM tensor and neglect the kinetic term contribution.

Our renormalization prescription in this section follows \cite{Onemli:2002hr} in the sense that we only introduce the dS invariant counter-terms. This is motivated to keep the equations of motion of the $ \lambda \phi^4$ theory intact. As in $D=4$, the dS symmetry will be broken by the renormalization. We will ask if the dS breaking counter-terms may or may not rescue the situation in the next section.


The EM tensor deriving from the potential term is
\beq
T^{{\rm mat}}_{\mu\nu}(x)_{\rm{pot.}}&=&- g_{\mu\nu}\left(\frac{\lambda+\delta\lambda}{4!} \phi^4 +\frac{1}{2} \delta m^2\phi^2 +\frac{\delta\Lambda}{8\pi G}\right)\nonumber\\
&\,& -2\delta\xi\left[g_{\mu\nu}((D-1)H^2 \phi^2  + (\phi ^2)^{; \rho}_{\,\,\, \rho})-(\phi^2)_{;\mu\nu}\right].\label{EMpot}
\eeq
To evaluate its expectation value, we expand the time-evolution operator as
\beq
\langle\Omega| T^{\rm mat}_{\mu\nu}(x)|\Omega\rangle\simeq\langle {T}\{T^{\rm mat}_{\mu\nu}(x)\left(1+i\int_C\sqrt{-g}d^2z \mathcal{L}_{\rm int}\right)\} \rangle,
\eeq
with $\mathcal{L}_{\rm int}$ made of the order $\lambda$ terms in \eqref{2dphi4}, in order to take into account the first order effects in the perturbation theory. The resulting expectation value of the EM tensor includes terms of order $\lambda^2$.
At the first order in $\lambda$, the expectation value of the EM tensor (\ref{EMpot}) can be evaluated in the free vacuum as we will see.

We begin with the evaluation of the following terms because we know they will determine the rest of (\ref{EMpot}).   
\beq
&&\hspace{-1cm} -g_{\mu\nu}\langle\Omega | \frac{\lambda}{4!}\phi ^4(x) +\frac{1}{2}\delta m^2 \phi^2(x)|\Omega \rangle\nonumber\\
&=&-g_{\mu\nu}\left[ \frac{\lambda}{4!}\langle\phi ^4(x)\rangle +\frac{1}{2}\delta m^2 \langle\phi^2(x)\rangle \right.
 \nonumber\\
&& +i\left(\frac{\lambda}{4!}\right)^2\int d^2z \sqrt{-g}\langle T\{ \phi ^4(x)\phi^4(z)\}\rangle 
- i\left(\frac{\lambda}{4!}\right)^2\int d^2z' \sqrt{-g}
\langle \phi^4(z')\phi ^4(x)\rangle 
\nonumber\\
&&
 + i\frac{\lambda}{4!}\frac{\delta m^2}{2}\int d^2z \sqrt{-g}
 \langle T\{\phi ^2(x)\phi^4(z)\}\rangle
- i\frac{\lambda}{4!}\frac{\delta m^2}{2}\int d^2z' \sqrt{-g}
\langle \phi^4(z')\phi ^2(x)\rangle
\nonumber\\
&&
+i\frac{\lambda}{4!}\frac{\delta m^2}{2}\int d^2z \sqrt{-g}
\langle T\{\phi ^4(x)\phi^2(z)\}\rangle
-i\frac{\lambda}{4!}\frac{\delta m^2}{2}\int d^2z' \sqrt{-g}
\langle \phi^2(z')\phi ^4(x)\rangle
\nonumber\\
&&
+i\left(\frac{\delta m^2}{2}\right)^2\int d^2z \sqrt{-g}
\langle T\{\phi ^2(x)\phi^2(z)\}\rangle
-i\left(\frac{\delta m^2}{2}\right)^2\int d^2z' \sqrt{-g}
\langle \phi^2(z')\phi ^2(x)\rangle
\nonumber\\
&&\left.
+\mathcal{O}(\lambda^3)\right].
\label{perturbation}
\eeq
We are working in the in-in formalism. The two copies of the vertices on so called $+$ and $-$ coordinates have been introduced: $z$ and $z'$ should be regarded as the vertices of the $+$ and $-$ types, respectively. The space-time point $x$ at which the EM tensor operator is inserted is now assumed to be on $+$ coordinate.

\subsection{Order $\lambda$ potential contributions}

\begin{figure}[tbp]
\begin{center}
\begin{minipage}{0.5\hsize}
\includegraphics[width=70mm]{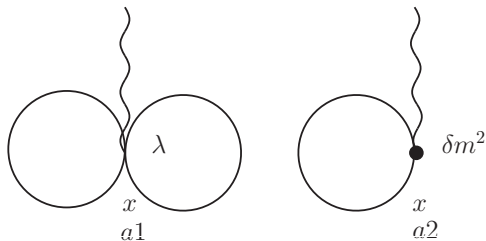}
\end{minipage}
\caption{Order $\lambda$ corrections to the EM tensor.}
\end{center}
\end{figure}

Let us consider the first two terms in (\ref{perturbation}) that are order $\lambda$ corrections to the matter EM tensor. The two diagrams corresponding to those terms are shown in Fig.1 $a1$ and $a2$. The first quantity to be calculated is the mass counter-term $\delta m^2$ which is determined by the renormalization condition that the renormalized mass is zero at the initial time $\eta_i=-1/H$ ($t_i=0$). This renormalization condition is the same as in \cite{Onemli:2002hr,Brunier:2004sb}. 
  
\begin{figure}[tbp]
\begin{center}
\includegraphics[width=80mm]{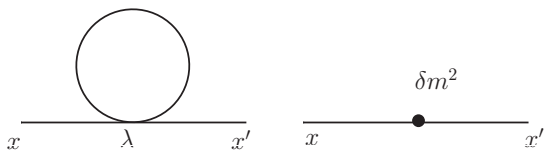}
\label{ct}
\caption{One-loop diagrams for mass corrections.}
\end{center}
\end{figure}
The one-loop diagrams shown in Fig. 2 give the one-loop scalar squared mass $M^2 _{\rm{1-loop}}$.
\beq
-iM^2_{\rm{1-loop}}(x,x')=-i\left[ \frac{\lambda}{2} \Delta(x,x)+\delta m^2\right] \delta^2(x-x').
\label{sqmass}
\eeq
 Substitution of (\ref{coincidence}) into (\ref{sqmass}) yields
\beq
M^2_{\rm{1-loop}}(x,x')=\left[ \frac{\lambda}{2}\alpha\beta \{ 2\ln(a(\eta))+\frac{2}{\omega}+A'\}+\delta m^2\right] \delta^2(x-x'),
\label{ssm}
\eeq
where $A'=C-\gamma$. The renormalization condition for the mass mentioned above reads
\beq
\delta m^2=-\frac{\lambda}{2} \alpha\beta\left( \frac{2}{\omega} +A' \right)+{O}(\lambda ^2).
\label{massc.t.}
\eeq
It follows that
\beq
M^2_{\rm{1-loop}}(x,x')&=&\frac{\lambda}{4\pi} \ln (a(\eta))\delta^2(x-x').
\eeq

To eliminate UV divergence from the EM tensor completely, we need to introduce the counter-term for the cosmological constant $\delta\Lambda$ in addition to the mass counter-term. The diagrams in Fig.1 $a1$ and $a2$ with the mass counter-term (\ref{massc.t.}) give 
\beq
\frac{\lambda}{4!} \langle\phi^4(x)\rangle+\frac{1}{2}\delta m^2 \langle\phi^2(x)\rangle &=&\frac{\lambda}{4!} \cdot 3\Delta^2 _{++}(x,x)+\frac{1}{2}\delta m^2 \Delta _{++}(x,x)\nonumber\\
&=&\frac{\lambda}{2}\left(\frac{1}{4\pi} \left(\frac{H^2}{4\pi}\right)^{\frac{\omega}{2}}\frac{\Gamma(1-\omega)}{\Gamma(1-\omega/2)}\right) ^2
\notag\\
&&\times  \left(-\frac{1}{\omega ^2}-\frac{1}{\omega }A'+\ln^2 a(\eta) -\frac{A'^2}{4}\right).
\eeq
Accordingly, $\delta \Lambda$ is determined by the requirement that
\beq
-g_{\mu\nu}\left[ \frac{\lambda}{4!}\langle\phi^4(x)\rangle+\frac{1}{2}\delta m^2 \langle\phi^2(x)\rangle\right]-\frac{\delta \Lambda}{8\pi G}g_{\mu\nu}=(\rm{finite}).
\eeq
Thus we have
\beq
\frac{\delta \Lambda}{8\pi G}= \frac{\lambda}{32\pi^2} \left(\frac{H^2}{4\pi}\right)^{-\omega}\frac{\Gamma^2(1-\omega)}{\Gamma^2(1-\omega/2)}\left[\frac{1}{\omega^2}+\frac{1}{\omega}A'\right]+\frac{\delta \Lambda_{\rm{fin}}}{8\pi G},
\eeq
where $\delta\Lambda_{\rm fin}$ is the finite part of the counter-term for which we shall choose $\delta\Lambda_{\rm fin}=A'^2/4$ at this stage. As a result, the EM tensor at order $\lambda$ is obtained as
\beq
\langle T^{{\rm mat}}_{\mu\nu}\rangle_{\rm{pot.}\,\, \mathcal{O}(\lambda)}&=&-g_{\mu\nu} \frac{\lambda}{32\pi^2}\ln ^2 a(\eta).
\label{orderlambda}
\eeq
As we have mentioned in the last subsection, the order $\lambda$ contribution corresponds to the zeroth order result in the perturbative expansion. In the next subsection and Appendix A, we consider the order $\lambda^2$ contribution in order to include the effect from the interaction vertices.
\subsection{Order $\lambda^2$ potential contributions}

In order to evaluate the renormalization of the cosmological constant at the $\lambda^2$ order, we must deal with the three-loop diagrams. 
We note that the counter-terms for the coupling constant $\delta \lambda$ and the conformal coupling $\delta \xi$ are absent in our computation. It is because the one-loop correction to the $\lambda \phi^4$ interaction term is not UV divergent in 2D and we do not have the terms proportional to the mixing of UV and IR divergent term $\omega^{-1}\cdot \ln a(\eta)$ that are supposed to be cancelled by $\delta \xi$ \cite{Onemli:2002hr}. The detail of calculation of the three loop diagrams is presented in Appendix A. The leading contribution to the EM tensor at this order is  
\beq
\langle T^{{\rm mat}}_{\mu\nu}\rangle_{\rm{pot.}\,\,\mathcal{O}(\lambda^2)}\sim -g_{\mu\nu}\frac{1}{8\pi}\frac{\lambda^2}{(4\pi)^2H^2} \ln^4 a(\eta).
\label{leading}
\eeq
The dimensionless expansion parameter can be regarded as $\lambda/H^2$ and our perturbative computation is valid as long as $\ln a(\eta)<H\lambda^{-1/2}$.

From (\ref{orderlambda}) and (\ref{leading}), we finally obtain the effective cosmological constant at the order $\lambda^2$,
\beq
\hspace{-0.5cm}\Lambda_{\rm{eff}}&=&\Lambda-
\frac{\kappa}{2}\langle T^{{\rm mat}\, \rho}_{\ \ \ \rho}\rangle
\notag\\
&\sim&  \Lambda +({\rm Weyl\ anomaly}) +\frac{\kappa\lambda}{32\pi^2}\ln ^2 a(\eta)+ \frac{1}{8\pi}\frac{\kappa\lambda^2}{(4\pi)^2H^2} \ln^4 a(\eta),
\eeq
where the Weyl anomaly is given by $\langle T^{{\rm mat}\, \rho}_{\ \ \ \rho}\rangle =R/(24\pi)$ \cite{Birrell:1982ix}. The effective cosmological constant has time dependence as  expected, and it increases as time passes. We recall that in the Liouville gravity the dS vacuum corresponds to negative value of $\Lambda$. Therefore, the cosmological constant evolves from the negative value toward zero, leading to the Minkowski space (within our approximation).
It means that the massless $\lambda \phi^4$ theory in $D=2$ shows the IR screening effect on the cosmological constant. As we noted in section 3, it crucially relies on the nature of the classical Liouville gravity. For comparison, see for instance \cite{Kitamoto:2011yx} where massless $\lambda \phi^4$ theory in 4D dS space has been investigated and the effective cosmological constant shows the anti-screening effect at the perturbative level.

\section{Do IR counter-terms recover the dS invariance? --- A brief look at Sine-Gordon model and $\lambda\phi^4$ theory} 

The conclusion that the $\lambda \phi^4$ theory screens the cosmological constant in late time is puzzling in $D=2$ dimensions. In flat Minkowski space, the IR limit of $\lambda \phi^4$ theory in $D=2$ dimensions is equivalent to a free Majorana fermion from the Landau-Ginzburg construction \cite{DiFrancesco:1997nk,Zamolodchikov:1986db}. The free Majorana fermion is conformally invariant and does not show any IR pathology in dS space.

A similar but slightly simpler question was addressed in the equivalence between Sine-Gordon model and massive Thirring model in the literature. In \cite{Bander:2010pn}, it was claimed that the equivalence is lost in dS space due to the quantum IR effects of the massless scalar propagator used in the perturbative construction of the Sine-Gordon model. However, as we presently show all the dS breaking effects are local and, if we allow the  {\it dS non-invariant} local counter-terms, these effects can be completely removed. Then the equivalence between Sine-Gordon model and the massive Thirring model still holds in dS space and both are dS invariant. 
It is worth asking if we could apply the same technique in $\lambda \phi^4$ theory to recover the dS invariance. In this paper, we give a brief report on this program and leave the details in a future publication.

We compare the $\lambda \phi^4$ theory with the Sine-Gordon model in dS background 
\begin{align}
S_{\text{SG}} = \frac{1}{2} \int d\eta dx \left( \partial_\eta \phi \partial_\eta \phi - \partial_x \phi \partial_x \phi - \frac{\lambda}{(H\eta)^2} \cos(\beta \phi) \right) \ . \label{gnaivesg}
\end{align}
We treat the sine-Gordon term ${\lambda}\cos(\beta \phi)$ as a perturbation around the free massless scalar theory similarly to the perturbative treatment of the $\lambda \phi^4$ theory we have discussed.

In perturbation theory with respect to $\lambda$, if we use the propagator with the IR cut off regularization
given by 
\begin{align}
G_{12} = \langle \phi(\eta_1,x_1) \phi(\eta_2,x_2) \rangle = -\frac{1}{4\pi}\ln\left( \frac{-(\eta_1 -\eta_2)^2 + (x_1 - x_2)^2}{H^{-2}} \right) \ , \label{prop}
\end{align}
the dS invariance is broken in Sine-Gordon model \cite{Bander:2010pn} precisely as in $\lambda \phi^4$ theory.
To obtain the dS invariance, the expression must be written by using the dS invariant length
\begin{align}
{y} = \frac{-(\eta_1 -\eta_2)^2 + (x_1 - x_2)^2}{\eta_1 \eta_2} \ , 
\end{align}
but then the simple replacement in the propagator such as 
\begin{align}
\mathcal{G}^{\text{inv}}_{12} = \langle \phi(\eta_1,x_1) \phi(\eta_2,x_2) \rangle = {-}\frac{{1}}{4\pi} \ln \left( \frac{-(\eta_1 -\eta_2)^2 + (x_1 - x_2)^2}{\eta_1 \eta_2} \right) \  \label{propr}
\end{align}
does not solve the massless equations of motion $\Box \phi = 0$ (except at $\eta=-\infty$).

We, however, realize the following alternative possibility.
If we modify the action with the time-dependent coupling constant
\begin{align}
S^{\text{modified}}_{\text{SG}} = \frac{1}{2} \int d\eta dx \left( \partial_\eta \phi \partial_\eta \phi - \partial_x \phi \partial_x \phi - \frac{\lambda (H\eta)^{\beta^2/4\pi}}{(H\eta)^{2}} \cos(\beta \phi) \right) \ , \label{gnaivesg}
\end{align}
then the de-Sitter invariance in correlation functions are recovered.
The time-dependent coupling constant precisely cancels the dS  non-invariant IR regularization in the massless scalar propagator (i.e. the IR logarithm). 
Therefore we may always use the IR counter-terms to remove {\it all} the dS breaking effects in Sine-Gordon model by declaring that \eqref{gnaivesg} is our dS invariant renormalized model.

With this viewpoint, let us reconsider the $\lambda \phi^4$ theory in 2D dS space.
The question is if we could remove all the dS breaking effects by introducing the IR counter-terms as in Sine-Gordon model. 
As for the cosmological constant, we could always cancel the dS breaking effects by introducing the {\it time-dependent} IR counter-terms of the cosmological constant  by $-\delta \Lambda(\eta) =  \frac{\kappa}{2} \langle T_{\rho}^{\ \rho}\rangle = \frac{\kappa\lambda}{32\pi^2}\ln ^2 a(\eta)+ \frac{1}{8\pi}\frac{\kappa\lambda^2}{(4\pi)^2H^2} \ln^4 a(\eta) + \cdots $. However, such cancellation seems ad hock.
Rather, the real question is if we could systematically set up the dS invariant perturbation theory by combining the dS breaking quantum IR effects in the propagator and the dS breaking time-dependent local counter-terms. If it is the case, we have a physical motivation to preserve the dS invariance at the sacrifice of the naive equations of motion.

Within the perturbative approach in $\lambda \phi^4$ theory, we find that the similar construction is non-obvious in contrast with the Sine-Gordon model. To illustrate this point, suppose that the dS invariant perturbation theory is achieved by adding the IR local counter-terms in our computation of the screening of the cosmological constant. 
At the first order in $\lambda$, we would expect the terms like $\lambda (\mathcal{G}^{\text{inv}}_{11})^2 = \lambda (G_{11} +\frac{1}{2\pi}\log (H\eta_1))^2$ (from diagram $a1$ of Fig.1). At this order, the necessary counter-term to reproduce the dS invariant VEV $\lambda (\mathcal{G}^{\text{inv}}_{11})^2$ is $\lambda \phi^2 \log (H\eta) $. At the second order in $\lambda$, we would like to  recover e.g. $\lambda^2 (\mathcal{G}^{\text{inv}}_{12})^4 = \lambda^2 (G_{12} +\frac{1}{4\pi}\log(H\eta_1)+\frac{1}{4\pi}\log(H \eta_2))^4$ (from diagram $b3$ of Fig.3 in Appendix A). However, in order to reproduce it from the {\it local} counter-terms, we would require the counter-terms $\lambda \phi^2 \log^2(H\eta) $ and $\lambda \phi^2$ among others. This is inconsistent with what we saw in the first order result. Thus, there is no local counter-terms to fully recover the dS invariant perturbation theory unlike in the Sine-Gordon example.\footnote{In the first version of the paper, we proposed that the curvature coupling may serve as the local counter-terms in all correlation functions, which turn out to be not the case. We would like to thank H.~Kitamoto for pointing out the inconsistency.}

\section{Discussion}

In this paper we have constructed a 2D model of quantum gravity coupled to matter in dS space to explore the quantum  IR effects in lower-dimensional dS space. The model (\ref{2Dmodel}) is described by the Liouville field theory coupled to matter which is minimally interacting with the Liouville field $\Phi$ through the physical metric. Once the fiducial metric is taken to be dS space, the classical Liouville field equation has a constant solution, and in this case the model reduces to an ordinary matter theory in the fixed dS background. One eminent feature of our model is that the cosmological constant in 2D dS space has the negative sign that follows from the Liouville field equation. This property of cosmological constant is opposite to the case of the Einstein gravity in $D>2$.

As a concrete matter Lagrangian, we have studied a massless scalar field theory with $\lambda \phi^4$ interaction minimally coupled to Liouville gravity. In dS space, the massless scalar propagator contains the IR divergence in the long wavelength limit and the IR logarithm appears due to the cutoff regularization of the IR divergence. Based on the in-in formalism, we have computed  the VEV of the EM tensor of order $\lambda^2$. The resulting VEV (\ref{leading}) has a time dependence through the IR logarithms, and as a consequence, the effective cosmological constant shows the screening effect at late time such that the absolute value decreases with time. This should be in contrast with the situations in $D>2$, in which the cosmological constant is anti-screened in the $\lambda \phi^4$ theory.

The degree of IR divergence in 2D, however, has turned out to be the same as that in 4D \cite{Onemli:2002hr}. If it were in Minkowski space, the degree of IR divergence in 2D would have been stronger than that in 4D. Nevertheless, the propagator in dS space is more complicated, and the structure varies by dimension. We do see the IR logarithms $\ln a$ both in 2D and 4D dS space, but we do not observe the enhanced degree of IR divergence in the VEV of the energy-momentum tensor.
 Based on this observation, we may expect that the same argument for the power-counting of the leading term of the IR logarithm in 4D dS space applies to our 2D case as well. According to \cite{Onemli:2002hr,Kitamoto:2011yx}, at $L$-loop order, the VEV of the energy-momentum tensor scales as: 
\beq
\langle T^{\rm mat}_{\mu\nu}\rangle\sim - g_{\mu\nu} \left( \frac{1}{H^2} \right)^{L-2} (\lambda \ln^2 a)^{L-1},\notag
\eeq
where the $L$ dependence of the power law of the Hubble constant compensates the mass dimension coming from the dimensionful coupling constant $\lambda$. Then  we may apply the known methods to resum the leading IR logarithms \cite{Starobinsky:1994bd,Burgess:2009bs} in our $D=2$ case, but we leave the detailed study for a separate work.

Once the effective cosmological constant is time dependent due to the matter quantum effects, the classical Liouville field dynamics will be affected as in the case of 4D Einstein gravity. The matter dynamics modifies the classical Liouville equation through the time-dependent matter EM tensor and it hinders  for the physical metric to possess the dS solution. 
We would like to investigate the dynamics of the subsequent Liouville field and its quantum effects in the back-reacted solution in a future work.

In order to claim that the observed dS breaking effects are physical, we have to ask if they may or may not be gotten rid of from the local counter-terms. Here, we should discuss rather unfamiliar time-dependent IR counter-terms. This possibility plays a crucial role to understand the (in)equivalence between Sine-Gordon model and massive Thirring model in dS space, where we have shown  that we are indeed able to recover the dS invariance by adding time-dependent IR counter-terms to the naive perturbative computations using the dS breaking propagator. Within the perturbation theory we have studied, however, we do not see that a similar mechanism works in $\lambda \phi^4$ theory. This fact supports the claim that the observed screening mechanism of the cosmological constant should be physical. It is desirable to establish the non-perturbative argument for the further support  because the perturbation becomes unreliable in later times with smaller effective cosmological constant. Further discussion on this sensitive issue will be found in our future publication.

\vskip7mm
\centerline{\bf Acknowledgments}\vskip1mm
We are grateful to Chong-Sun Chu, Tohru Eguchi, Kazuyuki Furuuchi, Hiroshi Isono, Katsushi Ito, Hiroyuki Kitamoto, Yutaka Matsuo, Yuji Tachikawa and Yuko Urakawa for valuable discussions. We are most thankful to Hiroyuki Kitamoto for pointing out a computational error in the section 5 in the original version of this paper. Correcting the error helped us to improve our understanding of the dS breaking IR effects. This work is supported in part by scientific grants from the Ministry of Education, Culture, Sports, Science and Technology under Grant Nos. 21244036 and 20012487 (T.Inami), National Science Council under Grant No. NSC-101-2112-M-007-021 and Taiwan String Theory Focus Group of NCTS under Grant No. NSC-103-2119-M-002- 001 (Y.Koyama), and Sherman Fairchild Senior Research Fellowship at California Institute of Technology and DOE grant number de-sc0011632 (Y.Nakayama).
%
\appendix

\section{Order $\lambda ^2$ corrections to the energy-momentum tensor}

In this appendix we outline the calculation of order $\lambda^2$ loop corrections to the EM tensor. From (\ref{perturbation}), we start with
\beq
&&\hspace{-1cm}-g_{\mu\nu}\langle \Omega | \frac{\lambda}{4!}\phi ^4(x) +\frac{1}{2}\delta m^2 \phi^2(x)|\Omega \rangle_{\mathcal{O} (\lambda^2)}\nonumber\\
&=&-g_{\mu\nu}\left[i\left(\frac{\lambda}{4!}\right)^2\int d^2z \sqrt{-g(z)}\langle T\{ \phi ^4(x)\phi^4(z)\}\rangle 
- i\left(\frac{\lambda}{4!}\right)^2\int d^2z' \sqrt{-g(z')}
\langle \phi^4(z')\phi ^4(x)\rangle 
\right.\nonumber\\
&& \quad
 + i\frac{\lambda}{4!}\frac{\delta m^2}{2}\int d^2z \sqrt{-g}
 \langle T\{\phi ^2(x)\phi^4(z)\}\rangle
- i\frac{\lambda}{4!}\frac{\delta m^2}{2}\int d^2z' \sqrt{-g}
\langle \phi^4(z')\phi ^2(x)\rangle
\nonumber\\
&&\quad 
+i\frac{\lambda}{4!}\frac{\delta m^2}{2}\int d^2z \sqrt{-g}
\langle T\{\phi ^4(x)\phi^2(z)\}\rangle
-i\frac{\lambda}{4!}\frac{\delta m^2}{2}\int d^2z' \sqrt{-g}
\langle \phi^2(z')\phi ^4(x)\rangle
\nonumber\\
&&\quad\left.
+i\left(\frac{\delta m^2}{2}\right)^2\int d^2z \sqrt{-g}
\langle T\{\phi ^2(x)\phi^2(z)\}\rangle
-i\left(\frac{\delta m^2}{2}\right)^2\int d^2z' \sqrt{-g}
\langle \phi^2(z')\phi ^2(x)\rangle\right].\notag\\
\label{lambda2}
\eeq
Note that we chose $x$ and $z$ as $+$ type vertices and $z'$ as $-$ type vertex as explained in section 4. The Feynman diagrams are shown in Fig.3 where the second line of (\ref{lambda2}) corresponds to $b1-b4$, the third line corresponds to $c1$ and $c2$, the fourth line corresponds to $d1$ and $d2$, and the last line corresponds to $e1$ and $e2$.
\begin{figure}[t]
\begin{center}
\begin{minipage}[b]{0.49\textwidth}
   \begin{center}
   \includegraphics[width=75mm]{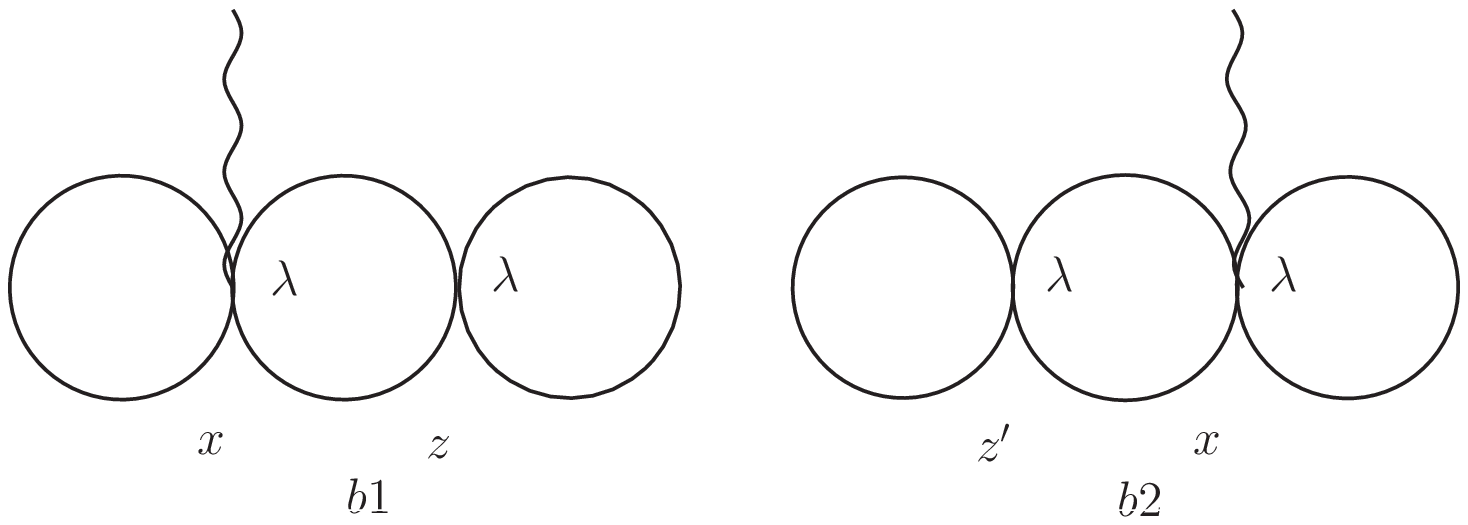}
   \end{center}
\end{minipage}   
\begin{minipage}[b]{0.49\textwidth}
   \begin{center}
   \includegraphics[width=50mm]{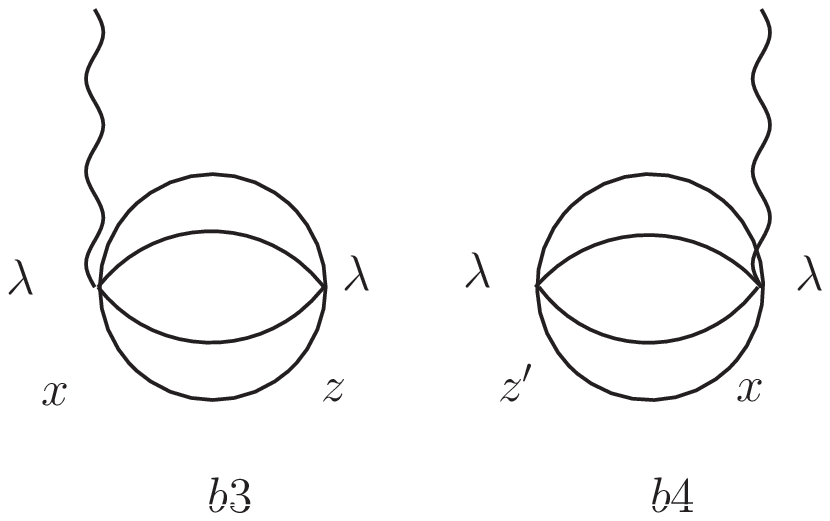}
   \end{center}
\end{minipage}   
\begin{minipage}[b]{0.49\textwidth}
   \begin{center}
   \includegraphics[width=70mm]{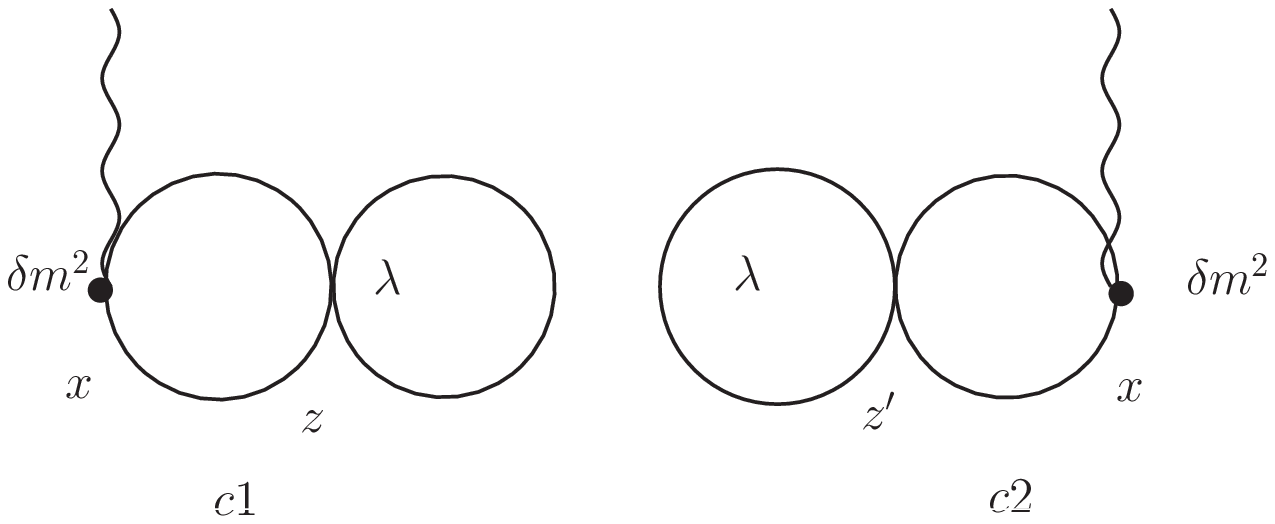}
   \end{center}
\end{minipage}  
 \begin{minipage}[b]{0.49\textwidth}
   \begin{center}
   \includegraphics[width=70mm]{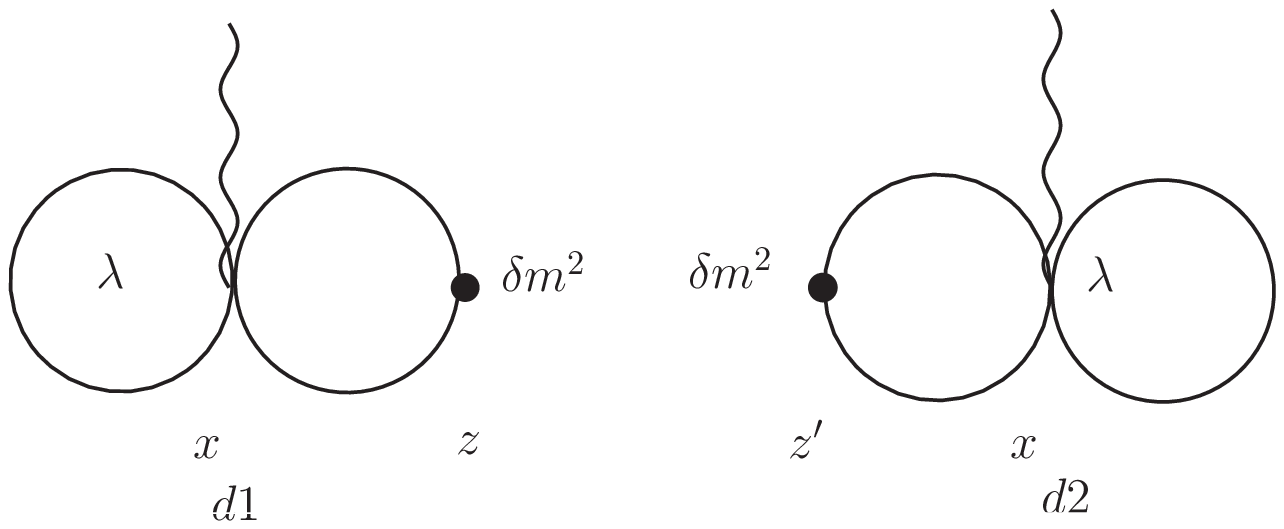}
   \end{center}
\end{minipage}   
\begin{minipage}[b]{0.49\textwidth}
   \begin{center}
   \includegraphics[width=65mm]{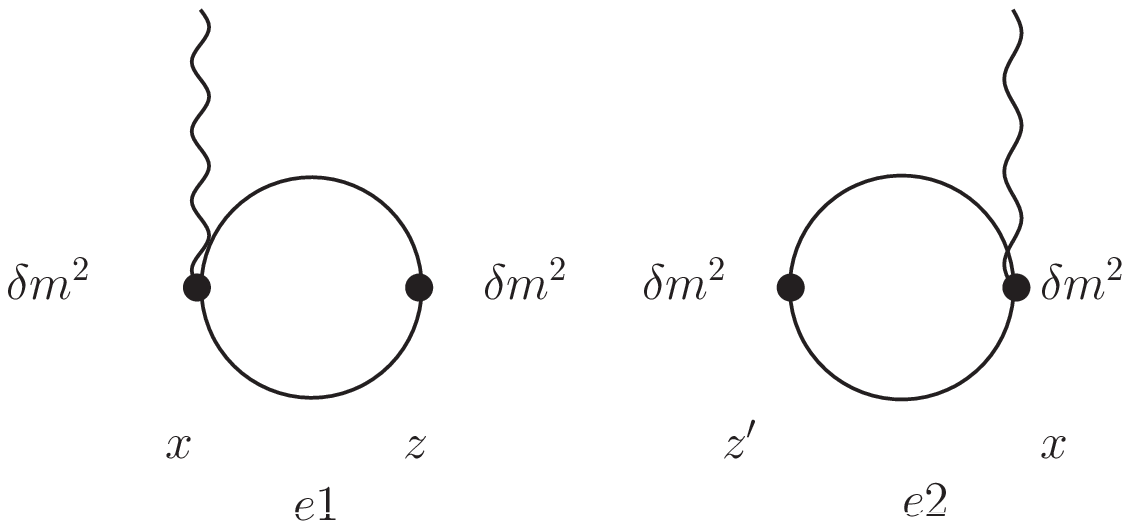}
   \end{center}
\end{minipage}
 \end{center}
\vskip-\lastskip
\caption{Order $\lambda^2$ corrections to the EM tensor $T_{\mu\nu {\rm pot}}$.}
\label{fig:label-3}
\end{figure}
Wick contractions and a simple calculation by the use of (\ref{propagators}) give
\beq
&&\hspace{-1cm}\langle \Omega | \frac{\lambda}{4!}\phi ^4(x) +\frac{1}{2}\delta m^2 \phi^2(x)|\Omega \rangle_{\mathcal{O} (\lambda^2)}\nonumber\\
&=&i\int d^2z\sqrt{-g}\left[ \frac{\lambda^2}{24} (i\Delta_{++}^4 (x,z)-i\Delta_{+-}^4 (x,z))\right.\nonumber\\
&&\left.
+\frac{1}{2} 
\left(\frac{\lambda}{2}i\Delta_{}(x,x)+\delta m^2\right) \left(\frac{\lambda}{2}i\Delta_{}(z,z)+\delta m^2\right)(i\Delta_{++}^2 (x,z)-i\Delta_{+-}^2 (x,z))\right]\notag\\
&\equiv& I_1+I_2,
\eeq
where we have collected the integrations into that of $z$ since $z$ and $z'$ are dummy variables, and defined $I_1$ and $I_2$ by
\beq
I_1(\eta)&\equiv& i\int d^2z\sqrt{-g}
\left[ \frac{\lambda^2}{24} (i\Delta_{++}^4 (x,z)-i\Delta_{+-}^4 (x,z))\right],\\
I_2(\eta)&\equiv& i\int d^2z\sqrt{-g}
\left[\frac{1}{2} 
\left(\frac{\lambda}{2}i\Delta_{}(x,x)+\delta m^2\right) \left(\frac{\lambda}{2}i\Delta_{}(z,z)+\delta m^2\right)(i\Delta_{++}^2 (x,z)-i\Delta_{+-}^2 (x,z))\right].\notag\\
\eeq

Let us first consider the $I_1(\eta)$. The integrand is expanded as 
\beq
&&\hspace{-1cm} i\Delta_{++}^4 (x,z)-i\Delta_{+-}^4 (x,z)\nonumber\\
 &=&\{ i\Delta_{++}^2 (x,z)+i\Delta_{+-}^2 (x,z)\}\{ i\Delta_{++} (x,z)+i\Delta_{+-} (x,z)\}\{ i\Delta_{++} (x,z)-i\Delta_{+-} (x,z)\}\nonumber\\
 &=&\alpha ^4 \left\{\gamma^4(y_{++} )-\gamma^4(y_{+-} )+4\beta \ln(aa_z)(\gamma^3(y_{++})-\gamma^3(y_{+-} ) )\right.\nonumber\\
 &\,&\left.+6\beta^2\ln ^2(aa_z)(\gamma^2(y_{++} )-\gamma^2(y_{+-} ))+4\beta^3\ln^3(aa_z)(\gamma(y_{++} )-\gamma(y_{+-} ))\right\},\label{subtract1}
\eeq
with abbreviations $y_{++}=y_{++}(x,z)$, $y_{+-}=y_{+-}(x,z)$, $a=a(\eta)$ and $a_z=a(\eta_z)$. It is clear from (\ref{pros}) that we can take a limit $\omega \to 0$ safely in (\ref{subtract1}) since the terms that include $\omega$ in their denominators offset each other. Thus we have  
\beq
&&\hspace{-1cm}i\Delta_{++}^4 (x,z)-i\Delta_{+-}^4 (x,z)\notag\\
&=& \left(\frac{1}{4\pi}\right)^4
 \left[\ln^4\left(\frac{y_{++}}{4}\right)-\ln^4\left(\frac{y_{+-}}{4}\right)
 -(4C+4\ln(aa_z))\left\{\ln^3\left(\frac{y_{++}}{4}\right)-\ln^3\left(\frac{y_{+-}}{4}\right)\right\}\right.\nonumber\\
&&\left.\qquad
+(6C^2-2\pi^2-36C\ln(aa_z)+6\ln^2(aa_z))
\left\{\ln\left(\frac{y_{++}}{4}\right)-\ln\left(\frac{y_{+-}}{4}\right)\right\}\right].\label{subtract2}
\eeq
Next we integrate (\ref{subtract2}) noting $\sqrt{-g}=a_z^2$ and
\beq
y_{++}(x,z)&=&aa_zH^2 (r^2-\Delta\eta^2+2ie|\Delta\eta|),\notag\\
y_{+-}(x,z)&=&aa_zH^2 (r^2-\Delta\eta^2-2ie\Delta\eta),
\eeq
with $r=|{\vec x}-{\vec z}|$ and $\Delta \eta=\eta-\eta_z$. The cut prescription allows us to write $\ln y$ as \cite{Onemli:2002hr,Kitamoto:2011yx}
\beq
\lim_{e\to 0}\ln y_{++}
&=& \ln\left[ aa_zH^2(\Delta\eta ^2 - r^2)\right] +i\pi \theta(\Delta\eta^2-r^2),\notag\\
\lim_{e\to 0}\ln y_{+-}
&=& \ln\left[aa_zH^2(\Delta\eta ^2 - r^2)\right] -i\pi \theta(\Delta\eta^2-r^2)(\theta(\Delta \eta)-\theta(-\Delta \eta)),\label{cut}
\eeq
 and then the interval of integration becomes
 \beq
\int d^2 z = \int^\eta _{-\frac{1}{H}} d\eta_z \int ^{\Delta \eta}_{0} dr.\label{integration}
\eeq
It means that the contribution from outside of the past light cone vanishes due to $i\Delta_{++}=i\Delta_{+-}$ for either $r^2>\Delta\eta^2$ or $\Delta\eta <0$. Using (\ref{cut}) and (\ref{integration}), we have
\beq
I_1(\eta)
&=&i \frac{\lambda^2}{24(4\pi)^4}\int ^\eta _{-\frac{1}{H}} d\eta_z a_z^2
 \Delta \eta \left[8\pi i \ln^3 (a a_z H^2 \Delta \eta ^2 /4) \right.\nonumber\\
&\,&\qquad+\{8\pi i K_1-6\pi i(4C+4\ln(a a_z))\}\ln^2 (a a_z H^2 \Delta \eta ^2 /4)\nonumber\\
&\,&\qquad+\{8\pi i (K_2-\pi ^2)-6\pi i K_5(4C+4\ln(aa_z))\nonumber\\
&\,&\qquad+4\pi i(6C^2-2\pi^2-36C\ln(aa_z)+6\ln^2(aa_z))\}\ln (a a_z H^2 \Delta \eta ^2 /4)\nonumber\\
&\,&\qquad+8\pi i (K_3-\pi^2 K_4)-(6\pi i K_6-2\pi^3 i)(4C+4\ln(aa_z))\nonumber\\
&\,&\qquad+4\pi i K_4\{6C^2-2\pi^2 -36C\ln(aa_z)+6\ln^2(aa_z)\} \nonumber\\
&\,&\qquad\left.+2\pi i\{-4C^3 -12\ln(aa_z)-12C\ln^2(aa_z)-4\ln^3(aa_z)\}\right],
\label{1terms}
\eeq
where $K_n$ ($n=1,\dots, 6$) are some constants which are not important in the subsequent discussions. For the time integral, it is convenient to change the variable $\eta_z$ to $a_z$,
\beq
\int ^\eta _{-\frac{1}{H}}d\eta_z = \int^{a}_1 \frac{da_z}{Ha_z}.
\eeq
The result of integral \eqref{1terms} has a very long expression and we shall avoid to present the full expression here because we are interested in the late time (namely $\eta \ll -H^{-1}$) behavior of $I_1(\eta)$ (and $I_2(\eta)$). The leading contributions from $I_1$ at that time can be extracted by 
\beq
I_1(\eta)\sim \frac{\lambda^2}{24\pi(4\pi)^2 H^2}\ln^4 a.\label{I1}
\eeq

Next we move on to the evaluation of the $I_2(\eta)$. It can be done in a way similar to that applied for the $I_1(\eta)$ and gives a simple expression. 
\beq
\hspace{-1cm}
I_2(\eta)&=&\frac{i}{2}\frac{\lambda^2}{(4\pi)^4}
\ln a 
\int d^2z\,\ln a_z\nonumber\\
&&\times\left\{\ln^2 y_{++} - \ln^2 y_{+-} 
 -2 (\ln(aa_z)+2\ln 2 +C)(\ln y_{++}-\ln y_{+-})
 \right\}\nonumber\\
&=& - \frac{{2}\pi \lambda^2}{(4\pi)^4H^2}\ln a
\int ^a _1 da_z\,\ln a_z  \left(-2-C +2 \ln\left(a_z^{-1}-a^{-1}\right)\right)\left(a_z^{-1}-a^{-1}\right).
\eeq
The integral of $a_z$ gives the result for $I_2(\eta)$,
\beq
I_2(\eta)&=&-\frac{2\pi\lambda ^2}{(4\pi)^4H^2} 
\left[ -\frac{2}{3}\ln^4 a-\frac{C}{2}\ln^3 a-\left(\frac{\pi^2}{3}-C\right)\ln^2 a\right.\nonumber\\
&& 
\left.-2{\rm Li}_2(a^{-1})-{\rm Li}_3(a^{-1})+\frac{2+C}{a}+\frac{\pi^2}{3}+2{\rm Li}_3(1)-2-C\right]
\label{timeint},
\eeq
where ${\rm Li}_n(x)$ denotes the polylogarithm function which decays for small $x$.
In this case, the leading contributions to the EM tensor is given by 
\beq
I_2(\eta)\sim \frac{\lambda^2}{{12}\pi(4\pi)^2H^2} \ln^4 a.\label{I2}
\eeq
Then the total contribution is given by (\ref{I1}) and (\ref{I2}),
\beq
I_1+I_2\sim \frac{1}{8\pi}\frac{\lambda^2}{(4\pi)^2H^2} \ln^4 a.
\eeq


\end{document}